\documentclass[brazil,english,aps,twocolumn,showpacs,prb]{revtex4}

\usepackage{amssymb}
\usepackage{pslatex}
\usepackage[T1]{fontenc}
\usepackage[latin1]{inputenc}
\usepackage{graphicx}
\usepackage{graphicx,epsf}
\usepackage{babel}
\usepackage{amsmath}



\begin{document}

\title{Fractional quantization of ballistic conductance in 1D
hole systems}
\author{M. Rosenau da Costa,$^{1}$ I.A. Shelykh,$^{1,2}$ and N.T. Bagraev$^{3}$}

\affiliation{$^{1}$International Center for Condensed Matter Physics, Universidade de Bras\'ilia, Caixa Postal 04667, 70910-900, Bras\'ilia-DF, Brazil\\
$^{2}$St. Petersburg State Polytechnical University, 195251, St. Petersburg, Russia\\
$^{3}$A.F. Ioffe Physico-Technical Institute of RAS, 194021, St.
Petesburg, Russia}

\begin{abstract}
We analyze the fractional quantization of the ballistic conductance
associated with the light and heavy holes bands in Si, Ge and GaAs
systems. It is shown that the formation of the localized hole state
in the region of the quantum point contact connecting two quasi-1D
hole leads modifies drastically the conductance pattern. Exchange
interaction between localized and propagating holes results in the
fractional quantization of the ballistic conductance different from
those in electronic systems. The value of the conductance at the
additional plateaux depends on the offset between the bands of the
light and heavy holes, $\Delta$, and the sign of the exchange
interaction constant. For $\Delta=0$ and ferromagnetic exchange
interaction, we observe additional plateaux around the values
$7e^{2}/4h$, $3e^{2}/h$ and $15e^{2}/4h$, while antiferromagnetic interaction plateaux are formed around $e^{2}/4h$, $%
e^{2}/h$ and $9e^{2}/4h$ . For large $\Delta$, the single
plateau is formed at $e^2/h$.
\end{abstract}

\pacs{73.23.-b, 73.21.Hb, 72.25.Mk}
\maketitle

The progress in nanotechnology allowed the fabrication of the quasi-
one dimensional mesoscopic components in which the transport of the
carriers has a ballistic character and is not accompanied by the
Joule losses \cite{Wees, Wharam}. The conductance through these
objects, $G$, is proportional to the number of the open 1D
propagation modes, $N$, which can be modulated by the application of
the perpendicular gate voltage, $V_g$. Tuning $V_g$ one observes
ballistic conductance staircase with plateaus at integer multiples
of the conductance quantum $G_{0}=2e^{2}/h$, where the factor 2
reflects the spin degeneration of the bands. This result can be
understood in  frameworks of the model of noninteracting electrons
\cite {Landauer-Buttiker}.

However, in the region of small carrier concentrations, an
additional plateau is observed near $0.7G_{0}$. In external magnetic
field it evolves smoothly into the spin-polarized conductance
plateau at $0.5G_{0}$ \cite{Thomas, Thomas2} thus indicating that
"0.7 feature" can be connected with spin. Later experiments  have
shown a zero-bias peak (ZBP) in the differential conductance and a
peculiar temperature dependence of "0.7 feature" \cite{Crone}, which
led to the proposal that it could be a Kondo-type phenomenon
\cite{Meir} connected with dynamical spin polarization. However,
more recent experiments demonstrated the absence of any ZBP at least
for some experimental configurations\cite {Grah} thus contradicting
the Kondo interpretation and supporting the models based on the
static spin polarization of the electrons in the region of the quantum point contact (QPC)
leading to the appearence of the spin gap \cite{Lassl,Rel,Bru}. The
formation of the spin polarized state in quantum wires and quantum
point contacts was predicted theoretically \cite{Wang,Bagraev,Rej2}
and obtained additional support from recent experimental studies
\cite{Grah,Rok}.

Within the model of the spontaneous spin polarization of the contact
the appearence of the "0.7 feature" can be explained as follows.
Suppose that QPC contains a single localized electron which affects
the propagating carriers via exchange interaction. Since the latter
is defined by the mutual orientation of their spins, the
transmission coefficient through the QPC with a magnetic moment is
spin-dependent. Besides, if the triplet state energy is lower than
the singlet state one, the potential barrier formed by the QPC
region for the carrier in the singlet configuration is higher than
for the triplet state. Therefore, at small concentration of carriers,
the ingoing electron in the triplet configuration passes the QPC
freely, while the carriers in the singlet configuration are
reflected, thereby defining the principal contribution of the
triplet pairs to the total conductivity. In zero magnetic fields the
probability of the realization of the triplet configuration is equal
to 3/4 against 1/4 for the singlet one, and thus the QPC conductance
in the regime considered reads $G=3/4G_0$  \cite{RejKuch}. On the
contrary, if the singlet configuration is energetically preferable,
the conductance should be equal to $G=1/4G_0$. The application of
the external magnetic field leads to the spin polarization of both
the propagating and localised carriers thus giving rise to the
conductance value equal to $G=1/2G_0$ in accordance with
experimental data. If more than one unpaired electrons is localized
on the QPC, the value of the conductance on the additional plateau
decreases \cite{Ivan}, which can explain the evolution of "0.7
feature" into "0.5 feature" with the increase of the length of the QPC
\cite{Reilly}.

In this work we analyze the ballistic conductance associated with
the holes bands in Si, Ge and GaAs  quasi 1D systems. We consider
the transmission of propagating hole states facing effective
potential barriers, generated by a spin-dependent interaction with a
localized hole, supposedly, present in the region of the QPC
\cite{Bird}. As it was mentioned above for electrons this model
qualitatively reproduces the major characteristics observed in the
0.7 experiments \cite{Ivan1}. Due to the difference of spin
structure for electrons and holes one can expect that the effect of
exchange interaction on ballistic conductance will be qualitatively
different in these two cases.

The valence bands of bulk Si, Ge and GaAs present two branches: the
heavy hole band, with spins $J_{z}^{hh}=\pm 3/2$, and the light hole band, with
spins $J_{z}^{lh}=\pm 1/2.$ In quasi-1D systems due to the effects of
confinement the energetic splitting $\Delta $ appears between these
two bands, which depends on the width of the quantum wire, the difference in
the effective masses of the light and heavy holes, $m_{lh}$ and
$m_{hh}$, and strains. In our further consideration we will take account of
only the lowest bands of the light and heavy holes \cite{strains}.
We suppose also that the propagating and localized holes interact
only in the region of the QPC having the length $L$. The Hamiltonian
of the system can be thus cast in the form:
\begin{equation}
H=\left\{
\begin{array}{c}
\frac{\hbar ^{2}k^{2}}{2m_{hh,lh}},\qquad x<0,x>L \\
\frac{\hbar ^{2}k^{2}}{2m_{hh,lh}}+V_{dir}+V_{ex}\mathbf{J}_{p}\mathbf{\cdot
J}_{l},\qquad x\in \left[ 0,L\right]
\end{array}
\right.   \label{4}
\end{equation}
where the indices $p$ and $l$ correspond to the propagating and
localized holes respectively, $V_{dir}>0$ is the matrix element of
the direct interaction and $V_{ex} $ is the matrix element of the
exchange interaction ($V_{ex}<0$ for ferromagnetic interacting and
$V_{ex}>0$ for antiferromagnetic interaction). It should be noted
that many-body correlations of the Kondo type can lead to the
temperature-dependent renormaliation of the exchange interaction
constant $V_{ex}$ \cite{Anderson} which can manifest itself in
non-trivial temperature dependence of the "0.7 feature" for both
electron \cite{Crone} and hole systems. The analysis of this
dependence, however, lies beyond the scope of the present work.

In the $\left\{ \left| m_{p},m_{l}\right\rangle \right\} $ basis the
Hamiltonian in the region of the QPC can be represented by a
block-diagonal $16\times 16$ matrix
\begin{equation}
H=diag\left[ H^{\left( +3\right) };H^{\left( +2\right) };H^{\left( +1\right)
};H^{\left( 0\right) };H^{\left( -1\right) };H^{\left( -2\right) };H^{\left(
-3\right) }\right] ,  \label{6}
\end{equation}
with each block associated with a given (superscript) value of the
z-component, $J_{T,z}=m_p+m_l$, of the total angular momentum of the pair of holes, $%
\mathbf{J}_{T}=\mathbf{J}_{p}+\mathbf{J}_{l}$:
\begin{widetext}
\begin{eqnarray}
H^{\left( \pm 2\right) } &=&\left(
\begin{array}{cc}
E_{hh}^{0}+\frac{3}{4}V_{ex}+\Delta  & \frac{3}{2}V_{ex} \\
\frac{3}{2}V_{ex} & E_{lh}^{0}+\frac{3}{4}V_{ex}+\Delta
\end{array}
\right) ,\quad H^{\left( \pm 1\right) }=\left(
\begin{array}{ccc}
E_{hh}^{0}-\frac{3}{4}V_{ex}+\Delta  & V_{ex}\sqrt{3} & 0 \\
V_{ex}\sqrt{3} & E_{lh}^{0}+\frac{1}{4}V_{ex}+2\Delta  & V_{ex}\sqrt{3} \\
0 & V_{ex}\sqrt{3} & E_{lh}^{0}-\frac{3}{4}V_{ex}+\Delta
\end{array}
\right) , \notag \\
H^{\left( 0\right) } &=&\left(
\begin{array}{cccc}
E_{hh}^{0}-\frac{9}{4}V_{ex} & \frac{3}{2}V_{ex} & 0 & 0 \\
\frac{3}{2}V_{ex} & E_{lh}^{0}-\frac{1}{4}V_{ex}+2\Delta  & 2V_{ex} & 0 \\
0 & 2V_{ex} & E_{lh}^{0}-\frac{1}{4}V_{ex}+2\Delta  &%
\frac{3}{2}V_{ex} \\
0 & 0 & \frac{3}{2}V_{ex} & E_{hh}^{0}-\frac{9}{4}V_{ex}
\end{array}
\right) ,\quad H^{\left( \pm 3\right) }=E_{hh}^{0}+\frac{9}{4}V_{ex},  \label{HT}
\end{eqnarray}
\end{widetext}
where $E_{hh,lh}^{0}=\frac{\hbar ^{2}k^{2}}{2m_{hh,lh}}+V_{dir}$.

The general expression for the conductance of the system at zero
temperature is given by
\begin{eqnarray}
G\left(E_F\right) &=&\frac{Ne^{2}}{h}\sum_{m_p
,m_l,m_p^{\prime},m_l^{\prime}=\pm 3/2,\pm 1/2}\alpha _{m_{p}}\left(E_F\right) \alpha _{m_{l}}\left(E_F\right)  \notag \\
&&\times \left| A\left(E_F\right) _{m_{p},m_{l}\rightarrow
m_{p}^{\prime },m_{l}^{\prime }}\right| ^{2}\delta
_{m_{p}+m_{l},m_{p}^{\prime }+m_{l}^{\prime }}.  \label{3}
\end{eqnarray}
where N is a number of the open propagating modes, $N=2$ if $%
E_{F}<\Delta$ and $N=4$ if $E_{F}>\Delta $, $\alpha
_{m_{p,l}}\left(E_F\right)$ are probabilities to find localized and
propagating hole in the state with spin projection $m_{p,l}$. In the
absence of an external magnetic field $\alpha _{\pm
3/2_{p,l}}\left(E_F\right)=1/2, \alpha _{\pm
1/2_{p,l}}\left(E_F\right)=0$ if $E_{F}<\Delta$ and $\alpha _{\pm
3/2_{p,l}}\left(E_F\right)=\alpha _{\pm
1/2_{p,l}}\left(E_F\right)=1/4$ if $E_{F}>\Delta$. The transmissions
amplitudes, $A_{m_{p},m_{l}\rightarrow m_{p}^{\prime },m_{l}^{\prime
}}$, are determined by finding the stationary states of the
corresponding propagating hole facing the effective potential
barrier described by the above $H^{\left( J_{T,z}\right) }$
matrices. Considering the conservation of the total spin,
represented by the Kronecker-$\delta $, we will get 44 different
transmissions amplitudes. In the absence of an external magnetic
field we will have the invariance as respect to the spin inversion,
and this number will be reduced to 22.

The transmissions amplitudes associated with the initial states $\left| +%
\frac{3}{2},+\frac{3}{2}\right\rangle $ and $\left| -\frac{3}{2},-\frac{3}{2}%
\right\rangle $ , corresponding to the values $J_{T,z}=\pm 3$, are
spin conservative and thus determined by solving the problem of a
free particle with kinetic energy $\hbar ^{2}k_{F}^{2}/2m_{hh}$
facing a square barrier of width $L$ and potential $V_{dir}+$
$\frac{9}{4}V_{ex}$. The first non-trivial spin-flip processes are
associated with the states with $J_{T,z}=\pm 2$. The transmission
amplitudes in this case can be determined from the procedure, which
we show for instance when the initial spin configuration of the
propagating and localized holes corresponds to the state $\left|
+\frac{3}{2},+\frac{1}{2}\right\rangle$.  The wavefunction of the
propagating hole reads

\begin{eqnarray}
&&
\begin{array}{c}
\Psi _{I}=\left(
\begin{array}{c}
1 \\
0
\end{array}
\right) e^{ik_{hh}x}+B_{\left| +\frac{3}{2},+\frac{1}{2}\right\rangle
\rightarrow \left| +\frac{3}{2},+\frac{1}{2}\right\rangle }\left(
\begin{array}{c}
1 \\
0
\end{array}
\right) e^{-ik_{hh}x} \\
+B_{\left| +\frac{3}{2},+\frac{1}{2}\right\rangle \rightarrow \left| +\frac{1%
}{2},+\frac{3}{2}\right\rangle }\left(
\begin{array}{c}
0 \\
1
\end{array}
\right) e^{-ik_{lh}x},\quad x<0,
\end{array}
\notag \\
&&
\begin{array}{c}
\Psi _{II}=\mathbf{X}_{1}\left(
C_{1}^{+}e^{ik_{1}x}+C_{1}^{-}e^{-ik_{1}x}\right) \\
+\mathbf{X}_{2}\left( C_{2}^{+}e^{ik_{2}x}+C_{2}^{-}e^{-ik_{2}x}\right)
,\quad x\in \left[ 0,L\right] , \label{WF}
\end{array}
\\
&&
\begin{array}{c}
\Psi _{III}=\left(
\begin{array}{c}
1 \\
0
\end{array}
\right) A_{\left| +\frac{3}{2},+\frac{1}{2}\right\rangle \rightarrow \left| +%
\frac{3}{2},+\frac{1}{2}\right\rangle }e^{ik_{hh}x} \notag \\
+\left(
\begin{array}{c}
0 \\
1
\end{array}
\right) A_{\left| +\frac{3}{2},+\frac{1}{2}\right\rangle \rightarrow \left| +%
\frac{1}{2},+\frac{3}{2}\right\rangle }e^{ik_{lh}x},\quad x>L,
\end{array}
\end{eqnarray}

where
\begin{eqnarray}
\left(
\begin{array}{c}
1 \\
0
\end{array}
\right)=\left|+\frac{3}{2}\right\rangle, \left(
\begin{array}{c}
0 \\
1
\end{array}
\right)=\left|+\frac{1}{2}\right\rangle,
k_{hh,lh}=\sqrt{\frac{2m_{hh,lh}E_{F}}{\hbar ^{2}}},
\end{eqnarray}
$\mathbf{X}_{1,2}$ are the eigenvectors of the matrix $H^{\left(
+2\right) }$ and the wave vectors $k_{1,2}$ are determined self
consistently in such a way that the eigenenergies, $\varepsilon
_{1,2}^{\left( +2\right) }\left( k_{1,2}\right) $, correspond to the
total energy of the system, $E_{F}+\Delta $ (once that in the
considered case we initially have  a propagating heavy hole, with
energy $E_{F},$ and a localized light hole, with energy $\Delta $).
The transmission and reflection amplitudes, $A$ and $B$ are found
from the continuity conditions at $x=0, x=L$. Analogous procedures
must be followed for all other possible initial states, grouping
them in the sets of conserved $J_{T,z}$ and diagonalizing
corresponding blocks of the Hamiltonian $H^{\left( \pm 1\right) }$
and $H^{\left( 0\right)}$.

\begin{figure}[tbp]
\includegraphics[width=1.0\columnwidth,keepaspectratio]{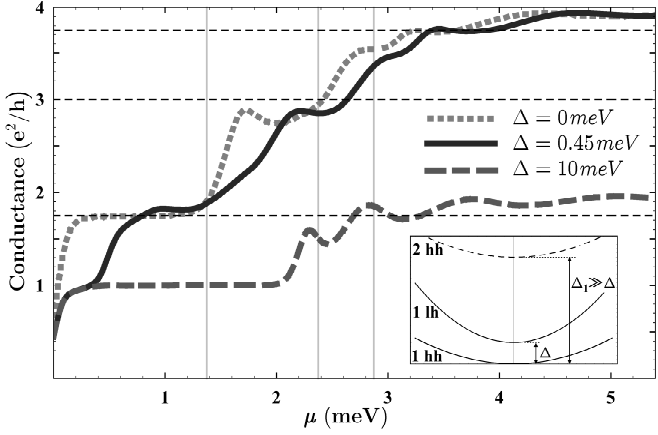}
\caption{Steps of the quantum conductance staircase vc the chemical
potential of carriers in a Si QPC for the ferromagnetic case. We considered
the standard values of the Si light and heavy holes effective masses: $%
m_{lh}=0.16m_{e}$, $m_{hh}=$ $0.49m_{e}$. The different lines correspond to
three different\ values of $\Delta (meV)=0,0.45,10$. The direct and exchange
interaction are estimated as $V_{ex}\simeq -0.5meV$ and $V_{dir}\simeq
1meV $, the length of the contact as $L=50nm$ and we considered the
temperature $T=0.5K.$ The vertical gray lines correspond to the values of the
heights of the effective potential barriers, $V_{dir}-3V_{ex}/4$, $%
V_{dir}-11V_{ex}/4$ and $V_{dir}-15V_{ex}/4$, whereas the dashed
horizontal lines correspond to the values $7e^{2}/4h$, $3e^{2}/h$
and $15e^{2}/4h$. The inset shows a band structure of the 1D holes.}
\label{Fig1}
\end{figure}

Figs.(1) and (2) show the conductance of a Si- based quantum well
for various offsets between the bands of light and heavy holes.
For ferromagnetic exchange interaction and moderate values of $%
\Delta $ we observe plateaux close to the values $7e^{2}/4h$,
$3e^{2}/h$ and $15e^{2}/4h,$ (Fig.(1)). For antiferromagnetic
interaction plateaux close to $e^{2}/4h$, $e^{2}/h$ and $9e^{2}/4h$
are formed (Fig.(2)). Considering the values of total spin of the
pair consisting of propagating and localized holes and counting the
corresponding probabilities of realization, we can show that these
values of conductance have a clear physical origin. The spin-
dependent part of the Hamiltonian (\ref{4}) can be rewritten as $V_{ex}%
\mathbf{J}_{p}\mathbf{\cdot J}_{l}=\frac{1}{2}V_{ex}\left( \mathbf{J}%
_{T}^{2}-\mathbf{J}_{p}^{2}-\mathbf{J}_{l}^{2}\right) =\frac{1}{2}V_{ex}%
\left[ J_{T}\left( J_{T}+1\right) -3\left( 3/2+1\right) \right] ,$
resulting in the values $9V_{ex}/4,$ $-3V_{ex}/4$, $-11V_{ex}/4$ and
$-15V_{ex}/4$, for the possible absolute values of the total spin
$J_{T}=3,2,1,0$, respectively. These values, summed with the direct
interaction, $V_{dir}$, will define four different spin dependent
heights of the effective potential barrier. The positive energy value of
these heights are plotted as vertical lines in
Figs.(1) and (2), and it is seen that they barely define the
energy limit of each plateaux. In the case of the ferromagnetic
interaction the barrier is lowest for the largest possible spin,
while in the case of the antiferromagnetic interaction the opposite
situation is realized. Now, for the unpolarized initial state, the
probabilities of
realization of the $J_{T}=3,2,1,0$ configurations of the total spin are $%
7/16 $, $5/16$, $3/16$ and $1/16$. Bearing in mind that in the case
$\Delta
=0$ the conductance of non-interacting system is quantized in units of $%
4e^{2}/h$, one easily obtains the values of the fractional conductance of  the
interacting system mentioned above.

\begin{figure}[tbp]
\includegraphics[width=1.0\columnwidth,keepaspectratio]{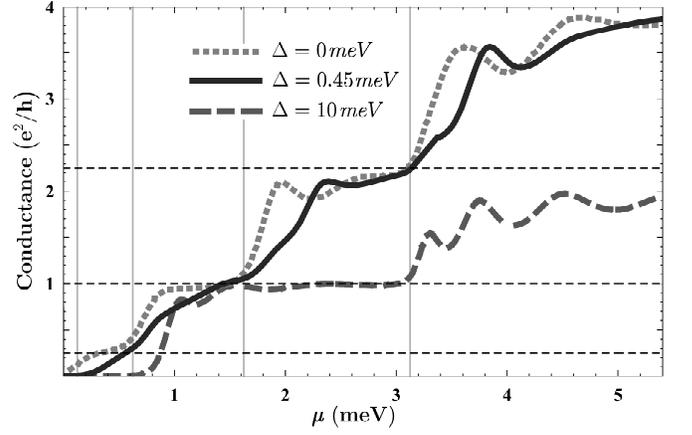}
\caption{Steps of the quantum conductance staircase vc the chemical
potential of carriers in a Si QPC for the antiferromagnetic case. We used
the same parameters of Fig.(1), only changing for an antiferromagnetic
exchange interaction, $V_{ex}\simeq 0.5meV$, and supposing $V_{dir}\simeq
2meV$.The vertical gray lines correspond to the values of the heights of the
effective potential barriers, $%
V_{dir}-15V_{ex}/4,V_{dir}-11V_{ex}/4,V_{dir}-3V_{ex}/4$ and $%
V_{dir}+9V_{ex}/4$, whereas the dashed horizontal lines correspond to the values $%
e^{2}/4h$, $e^{2}/h$ and $9e^{2}/4h$.} \label{Fig2}
\end{figure}

The situation is different if the offset between the bands of light
and heavy holes is not negligible. This version is illustrated  when $\Delta \rightarrow \infty $ that should be
realized in extremely narrow wires. In this case both localized and
propagating carriers are heavy holes, and thus the conductance of
non-interacting system is quantized in the units of $2e^2/h$. The spins
of the pair of holes can be either parallel or antiparallel.
Bearing in mind that in the absence of the external magnetic field
both configurations have equal probability and spin-flip processes
are blocked because of  the large offset of the intermediate light hole
states, one obtains just one additional plateau in the fractional
conductance corresponding  to $G= e^{2}/h$. The situation is thus
different from the case of electrons where spin-flip processes are
allowed and additional plateau is formed at $G=3e^{2}/2h$
\cite{RejKuch}.

When $\Delta $ is finite, it is clearly that the situation is beyond the
two considered extreme versions, as it is also illustrated in Figs.(1)
and (2). Both probabilities of realization of initial
states and transmission amplitudes are seen to be strongly energy
dependent.

The quantization of the ballistic conductance associated with holes in
Ge and GaAs structures is expected to be qualitatively the same as
in Si because of the similarity of the spin structure of the valence
band in these materials. However, it can be qualitatively different
in IV-VI semiconductors such as PbTe, PbSe and PbS where the
electron-hole symmetry holds.

\begin{figure}[tbp]
\includegraphics[width=1.0\columnwidth,keepaspectratio]{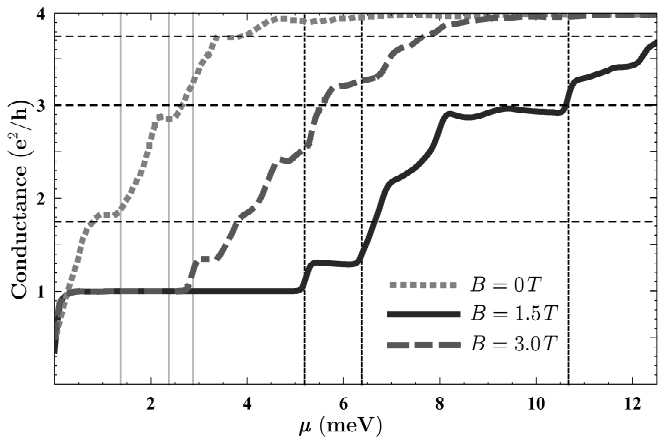}
\caption{Effect of an external magnetic field applied along the
growth axis of a Si QPC. Except for the applied magnetic field, we
used the same parameters of Fig.(1), with $\Delta =0.45meV$. We see
that moderate magnetic fields, $B=1.5T$, reduce the relative width
of the steps, change their position and increase their number, while
stronger ones, $B=3.0T,$ tend to reduce the effect of the spin
effective potential barrier, resulting in steps at integer values of
the quantum conductance. The values of the effective potential
barriers appear as vertical gray lines, while the energies of the
bottom of the subbands, appear as vertical dotted lines. Observe
that there is an inversion between the light and heavy hole subbands
because of  the difference in their effective g-factor and the
strong magnetic field.} \label{Fig3}
\end{figure}

In addition, we also analyzed the effects of an applied external
magnetic field parallel to the structure growth axis which produces
the Zeeman splitting of the light and heavy hole bands,
$H^{lh,hh}=-g_{\parallel }^{lh,hh}\mu _{b}J_{z}^{lh,hh}B\mathbf{,}$
where $\mu _{b}$ is the Bohr magneton and $g_{\parallel }^{lh,hh}$
are the parallel components of the effective g-factors tensors of
the light and heavy holes subbands. In our calculations for Si QPCs,
we estimate the g-factors as the ones of the bulk Si, $g_{\parallel
}^{lh}=2\kappa $ and $g_{\parallel }^{hh}=6\kappa $, with the
Luttinger parameter $\kappa =-3.41$ \cite{Hensel}. So, the propagating
and
localized light and heavy hole subbands will be split with the energies $%
\Delta _{Z}^{lh}=2\kappa \mu _{b}B$ and $\Delta _{Z}^{hh}=18\kappa \mu _{b}B$%
, respectively. In this case we do not have invariance as respect to
the spin inversion and the conductance of the system will be given
by the sum of all distinct 44 transmissions amplitudes present in
eq.(\ref{6}). The results of our calculation are summarized in
Fig.(3). The Zeeman splitting in each of the light and heavy hole
subbands, in addition to the $\Delta $ splitting between them,
increases the energy dependence of the initial state's probabilities
of realization and of the scattering processes. Therefore, under a
moderate magnetic field, the relative width of the steps in the
quantum conductance tends to be reduced, increasing in number and
changing in energy. This behavior is illustrated by the plotted
conductance for $B=1.5T$ in Fig.(3). However, if the external
magnetic field is strong enough such that the heights of the spin
effective potential barrier are smaller than the splitting between
the different subbands, then the conductance of the system tends to
show plateaux at the integer values of $e^{2}/h$. In the case of
$B=3.0T, $ in Fig.(3), the plateaux at $e^{2}/h$ and $3e^{2}/h$ are
very clear, while the plateau at $2e^{2}/h$ is absent, because the
spliting between the second and third subbands is smaller than the
heights of the spin effective potential barriers.

In conclusion, we have considered the effect of the exchange
interaction on the fractional quantization of the ballistic conductance
of the hole systems. It results in the fractional quantization of
the ballistic conductance different from those in electronic
systems. The value of the conductance at the additional plateaux
depends on the offset between the bands of the light and heavy holes
$\Delta$, the sign of the exchange interaction constant and external
magnetic field.

The authors thank A.C.F. Seridonio and V.L. Campo for productive
discussions. MRC acknowledge the FINEP and MCT for the financial
support. IAS and NTB acknowledge the support of the grants of the
President of Russian Federation and SNSF, Grant IB7320-110970/1.


\begin{references}

\bibitem{Wees}  B.J. van Wees \textit{et al.}, \textit{Phys. Rev. Lett.} \textbf{60}, 848
(1988).

\bibitem{Wharam}  D.A. Wharam \textit{et al.}, \textit{J. Phys. C} 21, L209 (1988).

\bibitem{Landauer-Buttiker}  R. Landauer, \textit{IBM J. Res. Dev.} \textbf{1}, 233 (1957);
M. Buttiker. \textit{Phys. Rev. Lett.} \textbf{57}, 1761 (1986)..

\bibitem{Thomas}  K.J. Thomas \textit{et al.} \textit{Phys. Rev. Lett. }\textbf{77}, 135
(1996).

\bibitem{Thomas2}  K.J. Thomas \textit{et al.} \textit{Phys. Rev. B.} \textbf{58}, 4846 (1998).

\bibitem{Crone}  S.M. Cronenwett \textit{et al.} \textit{Phys. Rev. Lett. }\textbf{88},
226805 (2002).

\bibitem{Meir}  Y. Meir et al, \textit{Phys. Rev. Lett.} \textbf{89}, 196802 (2002).

\bibitem{Grah}  A.C. Graham \textit{et al.}, \textit{Phys. Rev. B }\textbf{75}, 035331 (2007).

\bibitem{Lassl}  A. Lassl \textit{et al.} \textit{Phys. Rev. B}
\textbf{75}, 045346 (2007).

\bibitem{Rel}  D.J. Reilly, \textit{Phys. Rev. B }\textbf{\ 72}, 033309 (2005).

\bibitem{Bru}  H. Bruus \textit{et al.} \textit{Physica E}, \textbf{%
10}, (2001).

\bibitem{Wang}  Chuan-Kui Wang and K. F. Berggren, \textit{Phys. Rev. B }\textbf{54},
R14257 (1996). A.A. Starikov \textit{et al.}, \textit{Phys. Rev. B
}\textbf{67}, 235319 (2003);  P. Jaksch \textit{et al. Phys. Rev. B}
\textbf{74}, 235320 (2006)

\bibitem{Bagraev} N.T. Bagraev \textit{et al.} \textit{Phys. Rev. B} \textbf{70}, 155315 (2004).

\bibitem{Rej2}  T. Rejec and Y. Meir, \textit{Nature} \textbf{442}, 900 (2006).

\bibitem{Rok}  L.P. Rokhinson \textit{et al}. \textit{Phys. Rev.
Lett.} \textbf{96}, 156602 (2006).



\bibitem{RejKuch}  T. Rejec et al. \textit{J. Phys.: Cond. Matt.} \textbf{12}, L233-L239
(2000); V.V. Flambaum, M.Yu. Kuchiev, \textit{Phys. Rev. B}
\textbf{61}, R7869 (2000).

\bibitem{Ivan}  I.A. Shelykh \textit{et al.} \textit{Phys. Rev. B} {\bf %
74}, 085322 (2006).

\bibitem{Reilly} D. J. Reilly \textit{et al}. \textit{Phys. Rev. B} \textbf{63}, 121311(R) (2001);

\bibitem{Bird}  The possibility of the localization of the carriers in the region of QPC
was recently analyzed in experimental work by Y. Yoon \textit{et
al.} arXiv:0705.3019 (2007)

\bibitem{Ivan1} I.A. Shelykh \textit{et al.} \textit{J. Phys.: Cond. Matt}. 19, 246207 (2007)

\bibitem {strains} It should be noted that second level of heavy holes 2hh
lies usually very close to the first level of the light holes 1lh and in
general can not be neglected. However, the presence of the strains
usually leads to the lowering of the 1lh, and thus the situation
when it lies closer to hh1 than to 2hh can be realised. See E.L.
Ivchenko, G.E. Pikus, Superlattices and other heterostructures, ISBN
3-540-62030-3, Springer, 2003

\bibitem{Anderson} P.W. Anderson, \textit{J. Phys. C} \textbf{3}, 2436 (1970)

\bibitem{Hensel}  J.C. Hensel and K. Suzuki, \textit{Phys. Rev. Lett.} {\bf 22}, 838
(1969).

\end{references}
\end{document}